\documentclass[twocolumn,twoside,slac_two]{revtex4}
\usepackage{graphicx}
\usepackage{fancyhdr}
\usepackage{graphics}
\usepackage{epstopdf}
\usepackage{textpos}
\begin{document}

%%%%%%%%%%%%%%%%%%%%%% WRITE THE TITLE HERE %%%%%%%%%%%%%%%%%%%
\title{\centering Gamma-ray Astronomy:\\Implications for Fundamental Physics}
%%%%%%%%%%%%%%%%%%%%%% WRITE THE AUTHOR HERE %%%%%%%%%%%%%%%%%

%%% Please insert your personal picture here!

\author{
\centering
\begin{center}
J. Rico
\end{center}}
\affiliation{\centering  Instituci\'o Catalana de Recerca i Estudis
  Avan{\c c}ats (ICREA)\\ \& \\ Institut de F\'{\i}sica d'Altes
  Energies (IFAE), Bellaterra (Barcelona),
  08193, Spain}
%%%%%%%%%%%%%%%%%%%%%% WRITE THE ABSTRACT HERE %%%%%%%%%%%%%%%%
\begin{abstract}
  Gamma-ray Astronomy studies cosmic accelerators through their
  electromagnetic radiation in the energy range between $\sim$100 MeV
  and $\sim$100 TeV. The present most sensitive observations in this
  energy band are performed, from space, by the Large Area Telescope
  onboard the Fermi satellite and, from Earth, by the Imaging Air
  Cherenkov Telescopes MAGIC, H.E.S.S. and VERITAS. These instruments
  have revolutionized the field of Gamma-ray Astronomy, discovering
  different populations of gamma-ray emitters and studying in detail
  the non-thermal astrophysical processes producing this high-energy
  radiation. The scientific objectives of these observatories include
  also questions of fundamental physics. With gamma-ray instruments we
  study the origin of Galactic cosmic rays, testing the hypothesis or
  whether they are mainly produced in supernova explosions. Also, we
  obtain the most sensitive measurement of the cosmic
  electron-positron spectrum between 20 GeV and 5 TeV. By observing
  the gamma-ray emission from sources at cosmological distances, we
  learn about the intensity and evolution of the extragalactic
  background light, and perform tests of Lorentz Invariance. Moreover,
  we can search for dark matter by looking for gamma-ray signals
  produced by its annihilation or decay in over-density sites. In this
  paper, we review the most recent results produced with the current
  generation of gamma-ray instruments in these fields of research.
\end{abstract}

\maketitle
\thispagestyle{fancy}


\section{INTRODUCTION}
Gamma rays are the most energetic form of electromagnetic
radiation. They are produced in the non-thermal processes happening in
the most violent cosmic environments. The main production mechanisms
are radiation and interaction of accelerated charged
particles, either electrons or protons. Accelerated electrons may
produce gamma rays in the presence of magnetic fields by synchrotron
emission, by bremsstrahlung in the presence of matter, or by (inverse)
Compton scattering off ambient photons (sometimes those produced by
synchrotron emission of the same electron population). On the other
hand, gamma rays are also produced in the decay of the neutral pions
resulting from the interaction of accelerated protons with the
interstellar matter. Therefore, by studying gamma rays we learn about
cosmic particle accelerators. Those include supernova
remnants (SNRs), binary systems with a neutron star or a black hole,
pulsars, pulsar wind nebulae, starburst galaxies and active galactic
nuclei (AGNs). For a recent review see, e.g.,
\cite{Aharonian2011}. In addition, Gamma-ray Astronomy can
shed light on fundamental questions of Physics:
\begin{itemize}
\item[\emph{i)}] By studying the gamma-ray
emission produced by SNRs we can determine if that is produced by
electrons or protons, and answer the question of whether or not SNRs
are the main site of Galactic cosmic ray (CR) acceleration. 
\item[\emph{ii)}]
Current gamma-ray observatories are the most sensitive instruments to
measure the flux of cosmic electrons in the energy range between 20
GeV and 5 TeV, of special interest because it has been recently
claimed to contain a component which cannot be accounted by
conventional models of CR propagation and interaction in the
Galaxy. 
\item[\emph{iii)}]
By studying the gamma-ray emission from sources at
cosmological distances we can indirectly measure the extragalactic
background light (EBL). This is the light emitted by all extragalactic
sources over the history of the universe, and measuring it provides
constraints to models of star formation, galaxy evolution and
Cosmology. 
\item[\emph{iv)}] Distant and variable gamma-ray sources offer the
possibility to test Lorentz Invariance by looking for an
energy dependence of the speed of light. 
\item[\emph{v)}] Finally, we can search
for the gamma rays produced by dark matter (DM) annihilation or
decay in over-density sites like the satellite galaxies, the Galactic
center or galaxy clusters.
\end{itemize}

In this paper we present the most recent results produced by the
current gamma-ray observatories on these questions. In
Section~\ref{sec:instruments} we briefly describe the
present generation of gamma-ray instruments, and then address each of
the above-mentioned topics from Sections~\ref{sec:cr} to
\ref{sec:dm}. Finally, we summarize and conclude in
Section~\ref{sec:conclusions}.

\section{GAMMA-RAY INSTRUMENTS}
\label{sec:instruments}
Gamma rays are currently most sensitively detected, from space, by the
Large Area Telescope (LAT)\footnote{http://www-glast.stanford.edu/}
onboard the Fermi satellite and, from Earth, by the Imaging Air
Cherenkov Telescopes (IACTs)
MAGIC\footnote{http://wwwmagic.mppmu.mpg.de/},
H.E.S.S.\footnote{http://www.mpi-hd.mpg.de/hfm/HESS/} and
VERITAS\footnote{http://veritas.sao.arizona.edu/}. On a time scale of
five years, a new, more sensitive, IACT system, the Cherenkov
Telescope Array (CTA)\footnote{http://www.cta-observatory.org/}, could
already be in partial operation.

Fermi-LAT is composed of an anti-coincidence shield plus a tracker and
a calorimeter, which allow almost background-free, highly efficient
detection of gamma rays in the energy range between 30 MeV and more than 300
GeV. Fermi-LAT has a wide field of view (FoV) of about $4\pi/5$ steradian and
a duty cycle close to 100$\%$. It normally works in survey mode,
covering the full sky every three hours. It became operational in
August 2008, and, after the first year of observations, data are
publicly available in quasi realtime.  The Fermi-LAT Collaboration has
recently released the 2-year catalog (\cite{FermiCatalog}), with
almost 1900 detected sources (its predecessor, EGRET, detected a total
of 270 sources in its 10-years lifetime, \cite{EGRETCatalog}), many of
which still remain unidentified. 

On the other hand, IACTs are sensitive to the energy range from
$\sim$100 GeV to $\sim$100 TeV. The typical FoV of IACTs is of few
(3-5) degrees in diameter, and they usually operate in pointing mode,
with a duty cycle of $\sim$10$\%$. They image the Cherenkov light
produced in the electromagnetic showers initiated by cosmic radiation
in our atmosphere. The main background affecting the observations of
gamma rays using this technique is the overwhelming flux of charged
CRs --about 100 times more abundant than gamma rays for intense
sources--, which is reduced through the analysis of image
properties. Using this technique, over a dozen sources were detected
at energies of hundreds of GeV in the 1990s with the first generation
of IACTs.  These first exploratory instruments were then replaced in
the 2000s by the current generation of facilities, which have
revolutionized the field: gamma-ray source catalogues list now about a
hundred of sources and several new populations 
have been established as gamma-ray emitters, including SNRs, pulsar
wind nebulae, radio galaxies and gamma-ray binary systems (for a
recent review see, e.g.\ \cite{Funk2011}). The next breakthrough in
the field of gamma-ray astrophysics is bound to happen when CTA comes
online a few years from now. This instrument will consist of two
arrays of up to 100 telescopes of different reflector sizes in the
southern and northern hemispheres. The CTA arrays will be ten times
more sensitive than all the currently running experiments, and will
extend further the lower and higher ends of the energy sensitive
range.

\section{ORIGIN OF GALACTIC COSMIC RAYS}
\label{sec:cr}
CRs are energetic ($\sim10^8$ to $\sim10^{21}$ eV) particles which
bombard the Earth almost isotropically, discovered 100 years ago by
V. \cite{Hess1912}. CR energy spectrum reveals a non-thermal origin:
the flux is well described by a power law with index -2.7 up to $\sim
3 \times10^{15}$ eV (a feature dubbed the ``knee'' and for which the
absolute flux is $\sim 10^{-13}$ particles per m$^2$~sr~GeV~s),
and index -3.0 up to $\sim 3 \times10^{18}$ eV. They are composed
mainly by protons and heavier nuclei, but also contain a
non-negligible fraction of electrons, gamma rays, neutrinos and
anti-particles.

Galactic magnetic fields are intense enough so that CRs up to, at
least, $10^{16}$ GeV are confined within the Galaxy, and therefore it
is usually assumed that CRs up to the ``knee'' have a Galactic
origin. The assumption is supported by the fact that several Galactic
CR accelerator candidates are known, including SNRs, pulsar wind
nebulae, binary systems, star forming regions or super-bubbles. Among
them, and based on energy budget and abundance arguments, the main
contributors are thought to be SNRs: the total energy released by a
supernova is of the order of $10^{51}$ ergs and they happen with an
approximate frequency of one every $\sim$50 years. This should be
enough to produce all observed Galactic CRs if about 10\% of the total
energy was used on their acceleration.  Moreover, shock acceleration
provides a mechanism by which charged particles may attain the
observed energies: after the supernova explosion the ejected material
sweeps the interstellar and circumstellar gas forming a shock, which
is able to efficiently accelerate charged particles. Irrespectively of
the nature of the accelerated particles, electrons or protons, this
process produces gamma rays. Electrons lose energy through synchrotron
producing X-rays and radio emission, and through bremsstrahlung and
inverse Compton producing gamma rays. Protons lose energy through
proton-proton interactions, producing gamma rays by the decay of the
subsequent neutral pions. An important feature of the latter process is that
protons and gamma rays both follow a power law spectrum with the same
spectral shape, and the maximum energy attained by the gamma rays is
$\sim10\%$ of that of the protons.

\begin{figure}
\includegraphics[width=80mm]{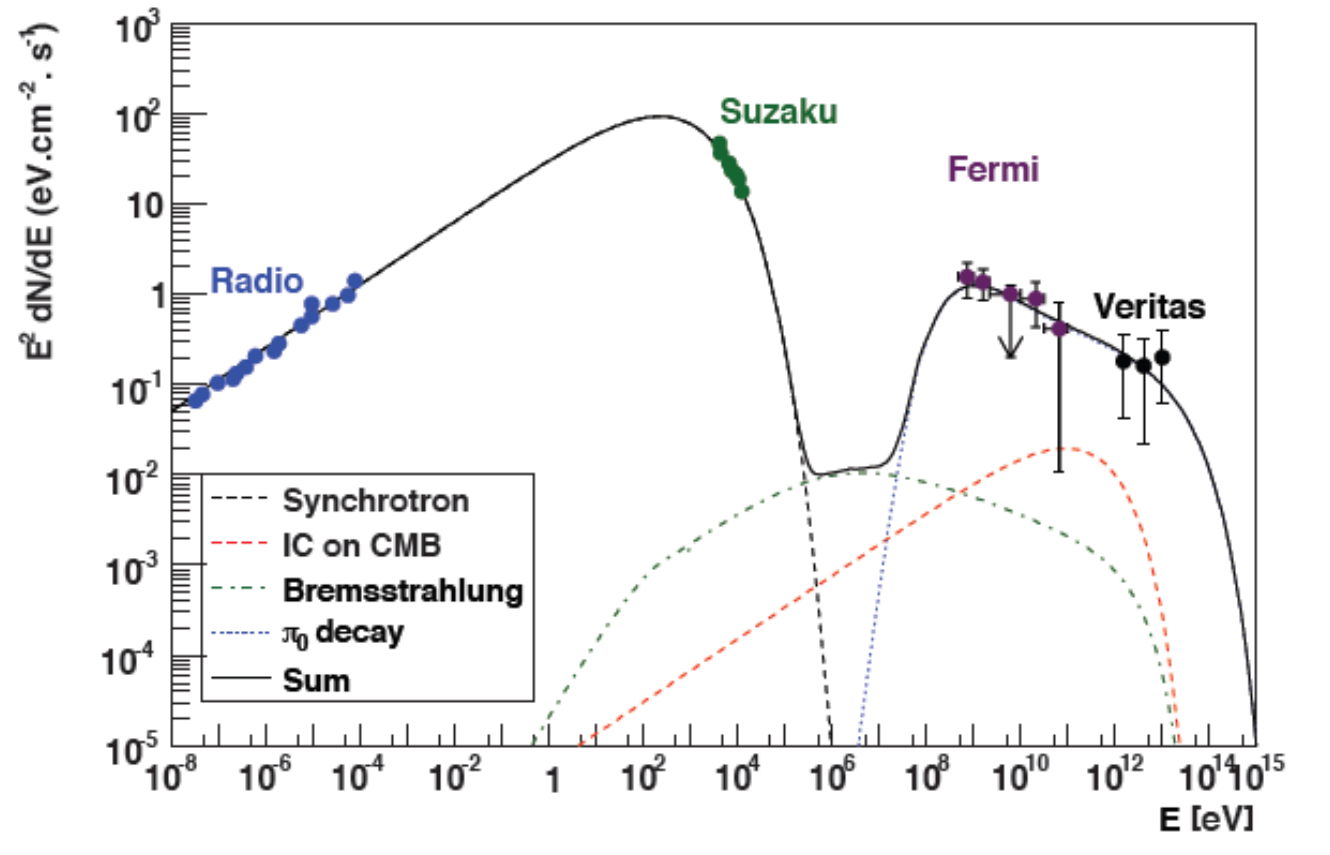}
\caption{Broadband SED model for Tycho SNR compared to experimental
  data. Figure taken from \cite{tychoFermi}.}
\label{fig:tycho}
\end{figure}

Gamma-ray emission is observed in several young shell-type SNRs. A
recent prototypical example is provided by the type Ia, shell-type SNR
Tycho. This source has an age of about 440 years, and lies at a
distance of 2-5 kpc. It has been detected in gamma rays by VERITAS
(\cite{tychoVeritas}) and Fermi-LAT (\cite{tychoFermi}). The magnetic
field is measured to be of 215~$\mu$G (\cite{tychoB}). This, together
with the synchrotron spectrum measured by radio and X-ray observations
constrains the electron population, whose spectrum can be well
described by a power-law with index $\sim-2.2$ and a cutoff at 6-7
TeV. In such a situation, the inverse Compton process of electrons off
photons of the Cosmic Microwave Background is not enough to reproduce
the observed gamma-ray flux, which would require a lower magnetic
field. Bremsstrahlung produces even lower fluxes at gamma-ray
energies. On the other hand, accelerated protons interacting with the
ambient medium are expected to produce gamma rays via pion decay, with
a spectrum that can be fitted to the observations (see
Figure~\ref{fig:tycho}). There are many examples of young shell-type
SNRs detected in gamma rays, like Cassiopeia A (\cite{CasAMAGIC},
\cite{CasAVeritas}, \cite{CasAFermi}), HESS~J1731-347
(\cite{1713Hess}), SN 1006 (\cite{1006Hess}), and others.

Another possible way to look for evidence of proton acceleration is to
search older SNRs close to dense molecular clouds. In this case the
number of targets for proton-proton collisions increases
dramatically and  gamma-ray emission is expected to be produced in
the interaction region between the remnant and the molecular cloud, or
by the ``illumination'' of farther clouds by the high energy protons
escaping the SNR shock. There are several examples of gamma-ray
sources compatible with this scenario, like IC443 (\cite{IC443Magic}, \cite{IC443Veritas},
\cite{IC443Fermi}), W51C (\cite{W51Hess},\cite{W51Fermi},
\cite{W51Magic}), W28 (\cite{W28Hess}, \cite{W28Fermi}) and others.

\begin{figure}
\includegraphics[width=38.6mm]{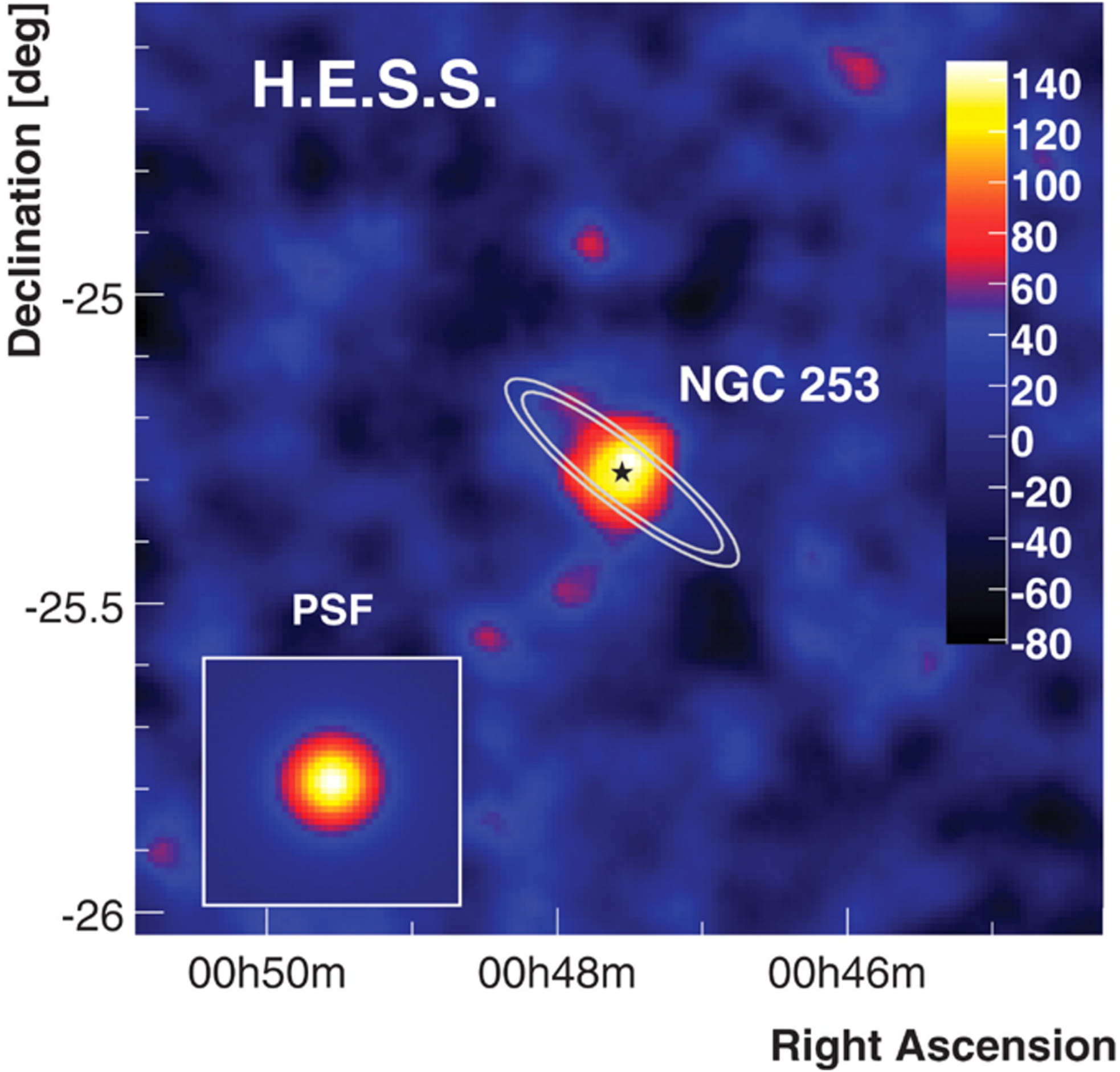}
\includegraphics[width=41.4mm]{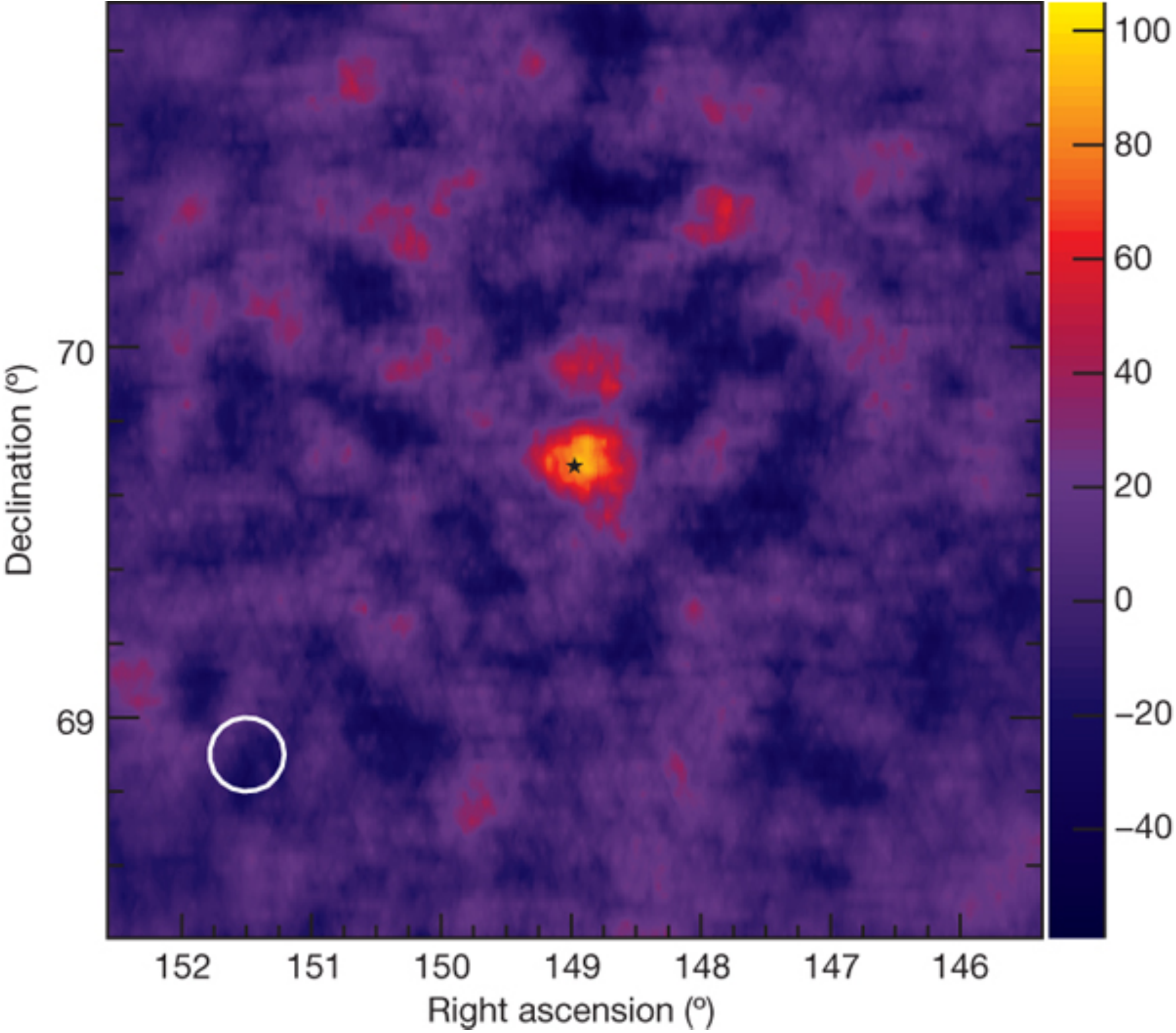}
\caption{Starburst galaxies detected in gamma rays. Left: NGC253
  as detected by H.E.S.S. after 119 hours of observation with a flux of
  0.3\% of the Crab Nebula above 220 GeV (\cite{NGCHess}). Right: M82 as detected by
  VERITAS after 137 hours of observation with a flux of 0.9\% of the
  Crab Nebula above 700 GeV (\cite{M82Veritas}).}
\label{fig:starburst}
\end{figure}

Further evidence for the Galactic origin of CRs up to the ``knee'' is
provided by the observations of starburst galaxies. This kind of
galaxies have large abundances of SNRs and massive star winds. If CRs
are accelerated by this kind of objects, starburst galaxies should
contain larger CR densities than our Galaxy, and we should detect the
gamma rays produced by their interaction with the interstellar gas
and radiation. The two most favorable cases have been recently
observed (see Figure~\ref{fig:starburst}). M82 has been
detected by VERITAS (\cite{M82Veritas}), with an inferred CR density
500 times larger than in our Galaxy. NGC253 has been detected by
H.E.S.S. (\cite{NGCHess}) with an estimated CR density up to 2000
times that of our Galaxy.  Fermi-LAT has also reported the detection
of gamma rays from these two starburst
galaxies (\cite{StarburstFermi}).

In summary, there is a growing evidence that SNR might be (one of) the
main accelerators of Galactic CRs, and some clear examples where such
acceleration is happening have been observed. However, there are still
some open questions. We still do not know if shocks are able to
accelerated particles up to the the highest Galactic CR energies. We
observe gamma rays from SNRs of up to $10^{13}$~eV, meaning that they
are produced by protons of up to $10^{14}$~eV, but we do not know if
energies up to the ``knee'' and beyond are reached or not. We still do
not know the ratio of protons and electrons accelerated in the shocks,
and if SNRs are the only Galactic CR accelerators. To solve these
questions, more observations allowing for a population study are
needed.

\section{COSMIC ELECTRON SPECTRUM}
\label{sec:electrons}

At the end of 2008, two independent measurements showing unexpected
features in the electron-positron (\cite{atic}) and positron fraction
(\cite{pamela}) spectra were reported. These discoveries immediately
triggered a burst of possible theoretical explanations, ranging from a
nearby pulsar to dark matter annihilation. Gamma-ray
instruments are contributing to shed light into this issue by providing the most
precise measurement of the electron-positron spectrum between 20 GeV
and 5 TeV, and of the excess in the positron
fraction between 20 and 200 GeV. 

\begin{figure}
\includegraphics[width=80mm]{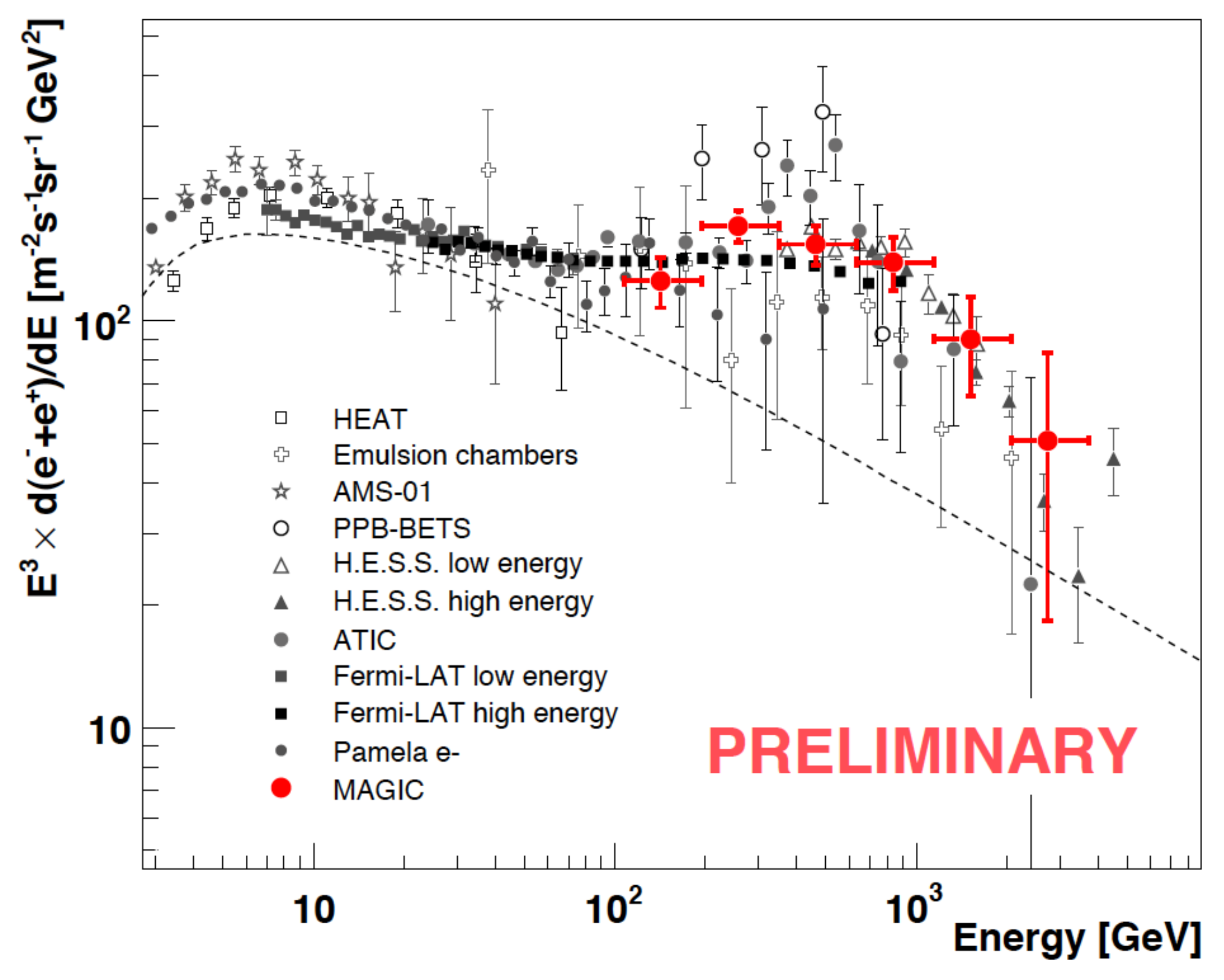}
\caption{Electron-positron spectrum in the energy range
  between 3 GeV and 5 TeV measured by different experiments. Figure
  taken from \cite{ElectronMagic}.}
\label{fig:electrons}
\end{figure}

Fermi-LAT is actually a fine electron spectrometer, and it records
$\sim 10^7$ electron-positrons per year above 20 GeV
(\cite{ElectronFermi}) . For this kind of measurement, however, it has
to deal with a non-negligible fraction of hadron
contamination. Separation of signal and background relies on extensive
Monte Carlo (MC) simulations validated with beam-test and flight
data. The measured spectrum is shown in
Figure~\ref{fig:electrons}. Fermi-LAT results are the most precise
between 20 GeV and 1 TeV. They show a smooth spectrum, and the
spectral feature claimed by previous experiments is not found. The
measurement deviates from the prediction of models of conventional CR
diffusion and interaction (\cite{galprop}), but it can be reasonably
well fitted by a $E^{-3.04}$ power law, when systematic errors are
considered.

H.E.S.S. and MAGIC have also measured the cosmic electron-positron
spectrum between 100 GeV and 5 TeV (\cite{ElectronHess},
\cite{ElectronMagic}). Electron and gamma-ray signals are in first
approximation indistinguishable for IACTs. The measurement is based on
the fact that the former are the dominant electromagnetic component of
the diffuse CR flux. The main background comes from hadronic CRs, and
can be reduced using multi-variate statistical analyses that exploit
the differences in the Cherenkov shower images between electrons and
hadrons (\cite{randomForest}), which requires extensive MC simulations of
shower development in the atmosphere. The results (see
Figure~\ref{fig:electrons}) are in agreement, within systematic
uncertainties, with what was observed by Fermi-LAT: the spectral
feature claimed by \cite{atic} is not confirmed, and the spectrum is
well fitted by a power law with index -3, and a steepening starting at
$\sim$1 TeV.

\begin{figure}
\includegraphics[width=80mm]{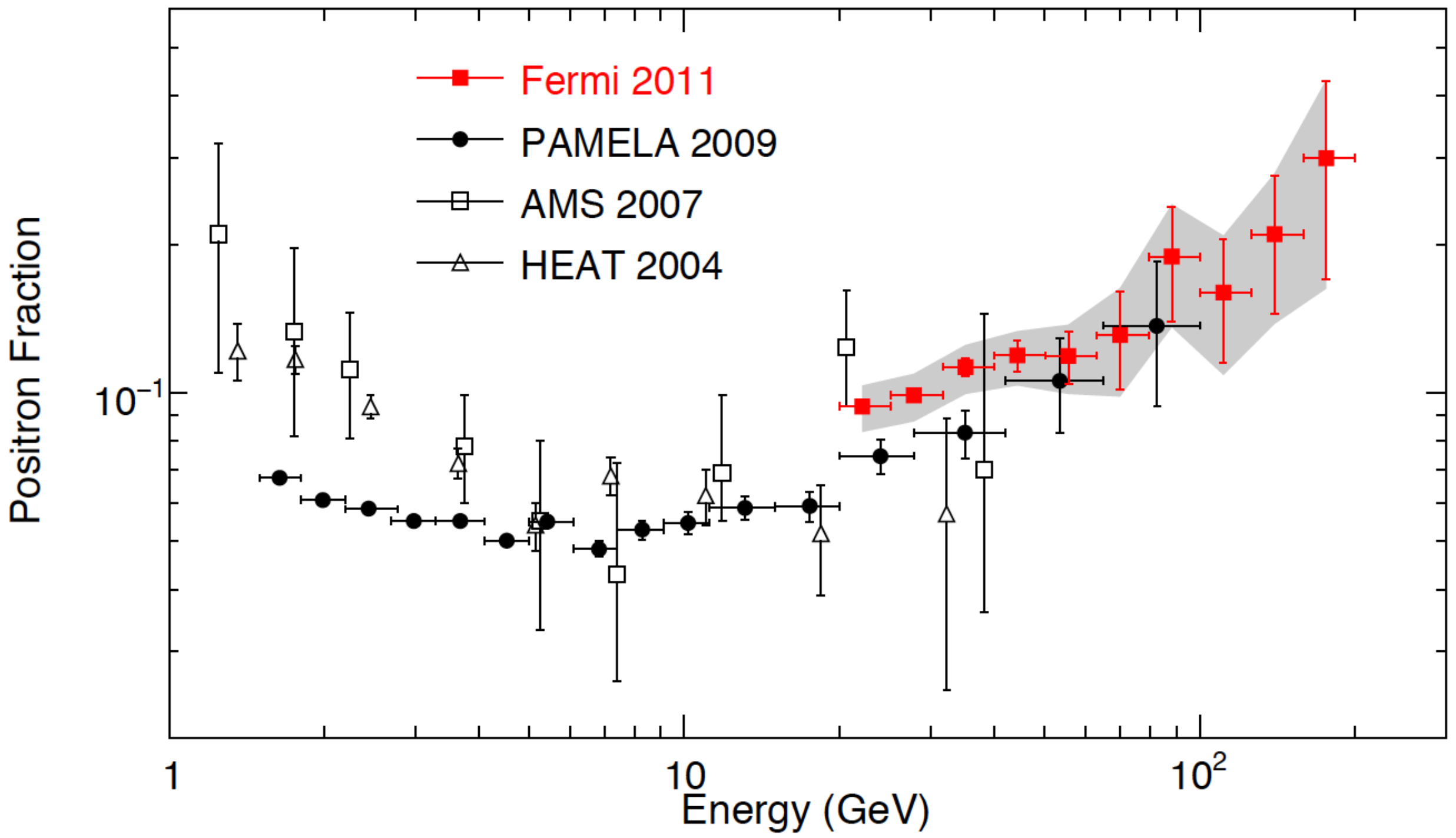}
\caption{Positron fraction measured by Fermi-LAT and other experiments
  in the energy range
  between 1 GeV and 200 GeV. Figure
  taken from \cite{PositronFermi}.}
\label{fig:positron}
\end{figure}

Fermi-LAT has also measured separate CR positron and electron spectra
(\cite{PositronFermi}). Since the detector has no magnet, both
populations are separated by exploiting the Earth's shadow on the
CR flux, whose position depends on the particle's charge. This
allows Fermi-LAT to produce a measurement of both components
separately, and of the positron fraction between 20 GeV and 200 GeV
(see Figure~\ref{fig:positron}). The results are in agreement with
\cite{pamela} and shows, for the first time, that the unexpected
increase in the positron fraction continues at least up to 200
GeV. 

MAGIC is currently developing a conceptually similar measurement that
exploits the Moon shadow and the dependence of its position on the
particle's charge (\cite{MoonMagic}). With this strategy, the
measurement of the positron fraction is expected to be extended up to
$\sim$700 GeV. This is a very challenging measurement due to
the high level of noise induced by the scattered moonlight during the
observations. Although the major technical drawbacks of the technique have
now been overcome, the  detection of the electron Moon shadow
with MAGIC will still require several years because of the
short observation window available every year.

\section{EXTRAGALACTIC BACKGROUND LIGHT}
\label{sec:ebl}

The EBL is composed of low-energy ($\lambda \sim 10^{-1}-10^{3} \mu$m)
photons radiated by stars and galaxies in the course of cosmic
history. The EBL spectrum contains information about the
history of the Universe, star formation, galaxy evolution and
cosmology. It consists of two main components: the redshifted light from
stars and the redshifted light reprocessed by dust. Although solid
lower limits to EBL can be obtained by galaxy count experiments,
direct measurements are extremely challenging, due to the  intense
foreground light, dominated by terrestrial, zodiacal and Galactic
sources, and which outshines the EBL over
the whole spectral range.

Gamma-ray instruments provide indirect constraints to the EBL at the
$\mu$m range. That is possible because gamma rays traveling
cosmological distances have a non-negligible probability of
interaction with EBL photons through $e^+e^-$ pair creation. That
probability depends on the gamma-ray energy, the traveled distance,
and the details of EBL (intensity, spectrum, time evolution,
etc). Therefore, the spectrum of cosmological gamma-ray sources
measured at Earth is the convolution of the \emph{intrinsic} spectrum
and the energy-dependent modification produced by the interaction with
the EBL. Thus, by measuring the gamma-ray spectrum of distant sources
and making some assumptions on the properties of the intrinsic one, we
can constrain the EBL.

\begin{figure}
\includegraphics[width=80mm]{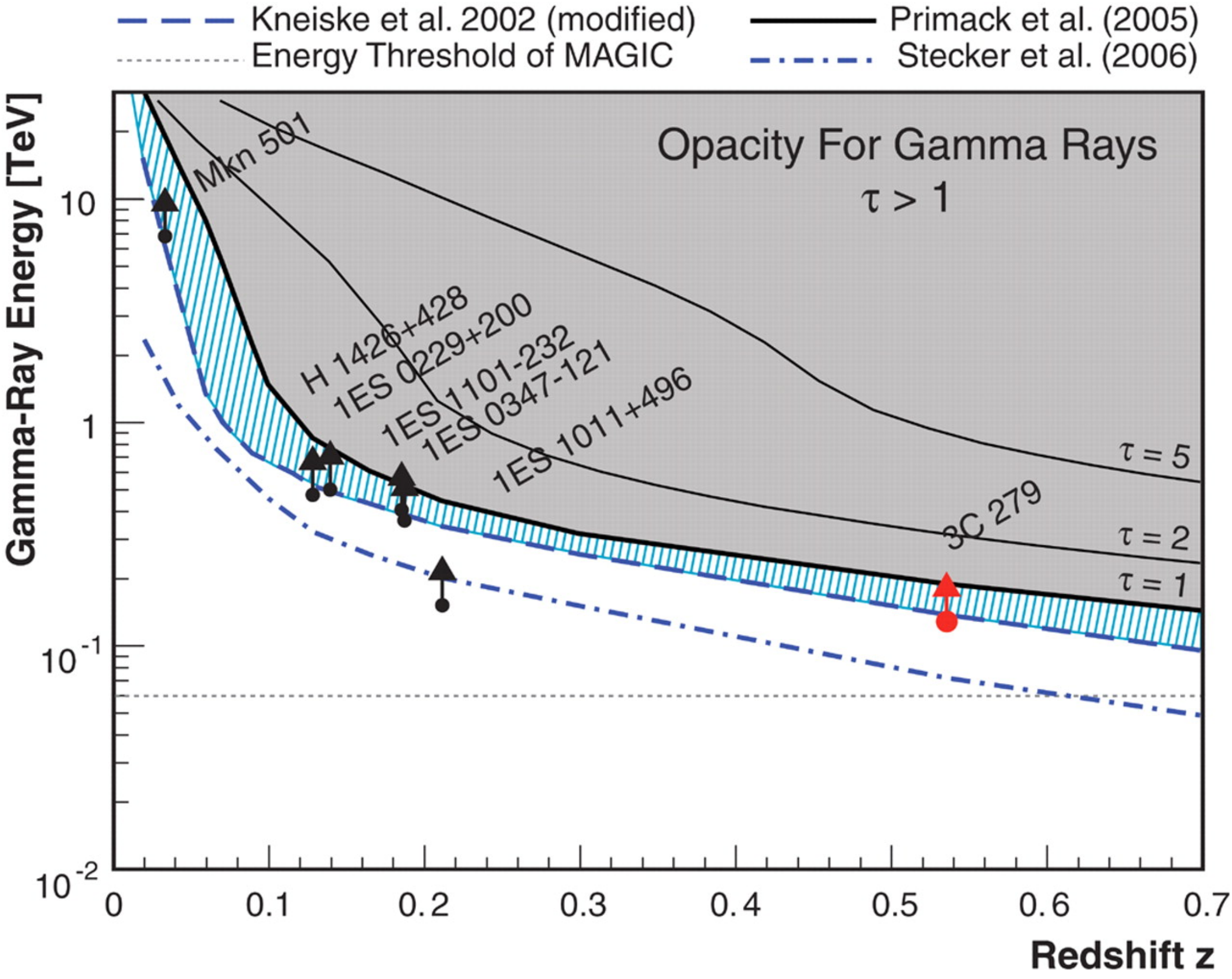}
\caption{For a given distance ($z$), the lines show, for different EBL
  models, the gamma-ray energy for which a reduction of the flux of
  $\sim$63\% is expected. Observation of AGNs at different distance
  and energies are used to constrain EBL models. Figure taken from
  \cite{EBLMagic}.}
\label{fig:EBLMagic}
\end{figure}

\begin{figure}
\includegraphics[width=80mm]{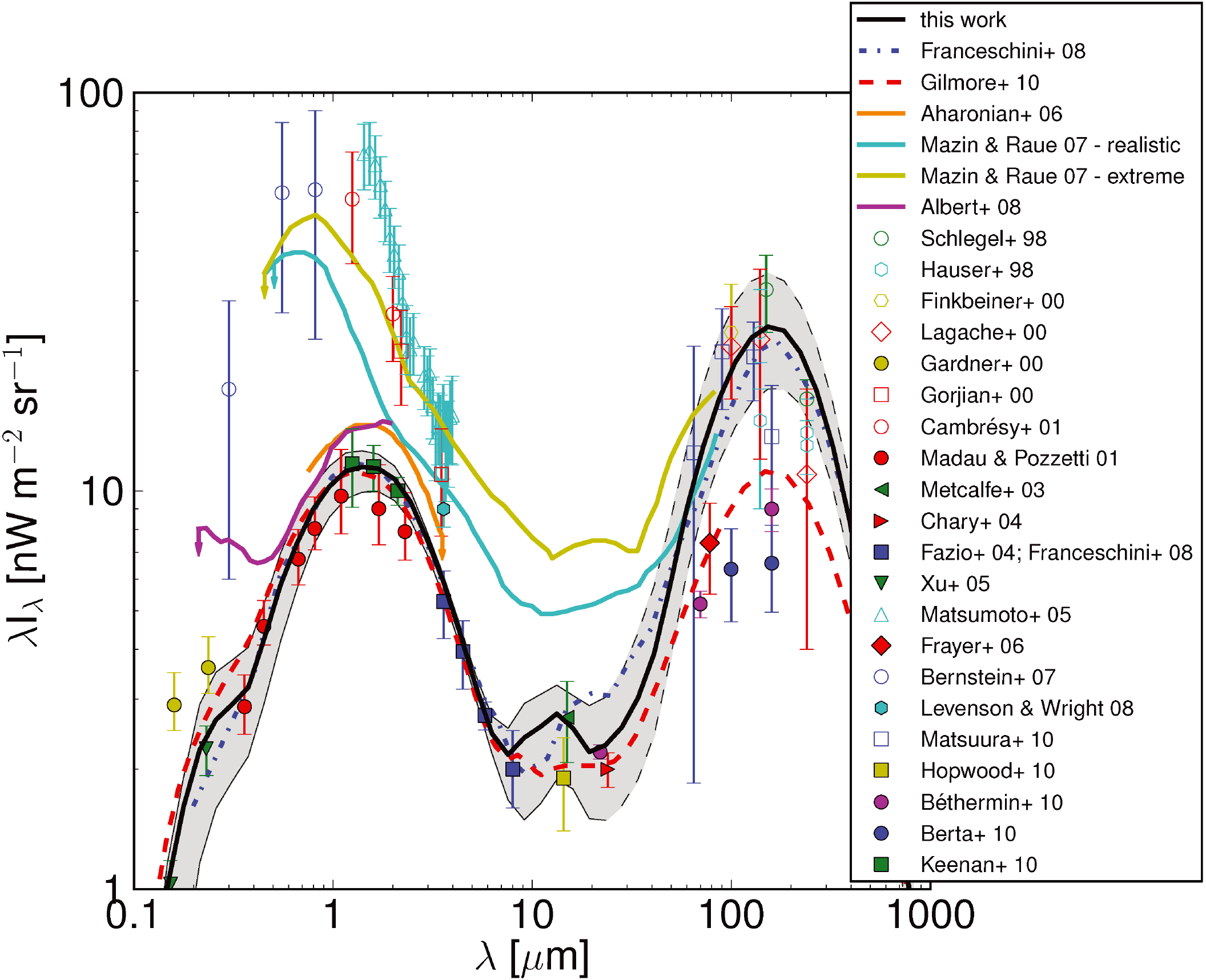}
\caption{Recent EBL limits, measurements and theoretical
  models. Figure taken from \cite{dominguez}.}
\label{fig:dominguez}
\end{figure}

H.E.S.S. (\cite{EBLHess}) and MAGIC (\cite{EBLMagic} and
Figure~\ref{fig:EBLMagic}) have set upper limits to the EBL using
observations of distant AGNs. In this case, two different assumptions
on the intrinsic spectrum are normally used, namely: that a fit using
a power-law yields a photon index larger than 1.5; and that there is
no spectral pile-up, i.e. no presence of an extra high-energy
component. Fermi-LAT (\cite{EBLFermi}), on the other hand, excludes
EBL models by imposing lower limits to the probability of detection of
the highest-energy photon recorded in observations of distant AGNs and
gamma-ray bursts (GRBs). These measurements have resulted in the
exclusion of all those models predicting high EBL levels, and present
limits and theoretical understanding suggest an EBL close to the lower
limits inferred from galaxy counts (see Figure~\ref{fig:dominguez} and
\cite{dominguez}).

In addition, if the EBL and the intrinsic spectrum of an observed
distant source are perfectly known, the measured spectrum depends only
on the distance to the source. Based on this idea, \cite{blanch} have
proposed to use the measured energy spectrum of distant sources to fit
cosmological parameters.

\section{TESTS OF LORENTZ INVARIANCE}
\label{sec:lorentz}

Gamma-ray instruments have also been used to test Lorentz
Invariance. Several Quantum Gravity theories predict that Lorentz
Invariance is not preserved at energy scales close to the Planck Mass
($M_P = 1.2\times 10^{19}$ GeV). In such a case, there should be an
energy dependence of the speed of light, whose expression can be
expanded as
\begin{equation}
v = c\left(1 \pm \xi\frac{E}{M_P} \pm \zeta^2 \left(
    \frac{E}{M_P}\right)^2 + ...\right)
\label{eq:v}
\end{equation}
where $\xi$ and $\zeta$ parameterize the strength of the linear and
quadratic dependences of the speed of light with the energy,
respectively.

\begin{figure}
\includegraphics[width=80mm]{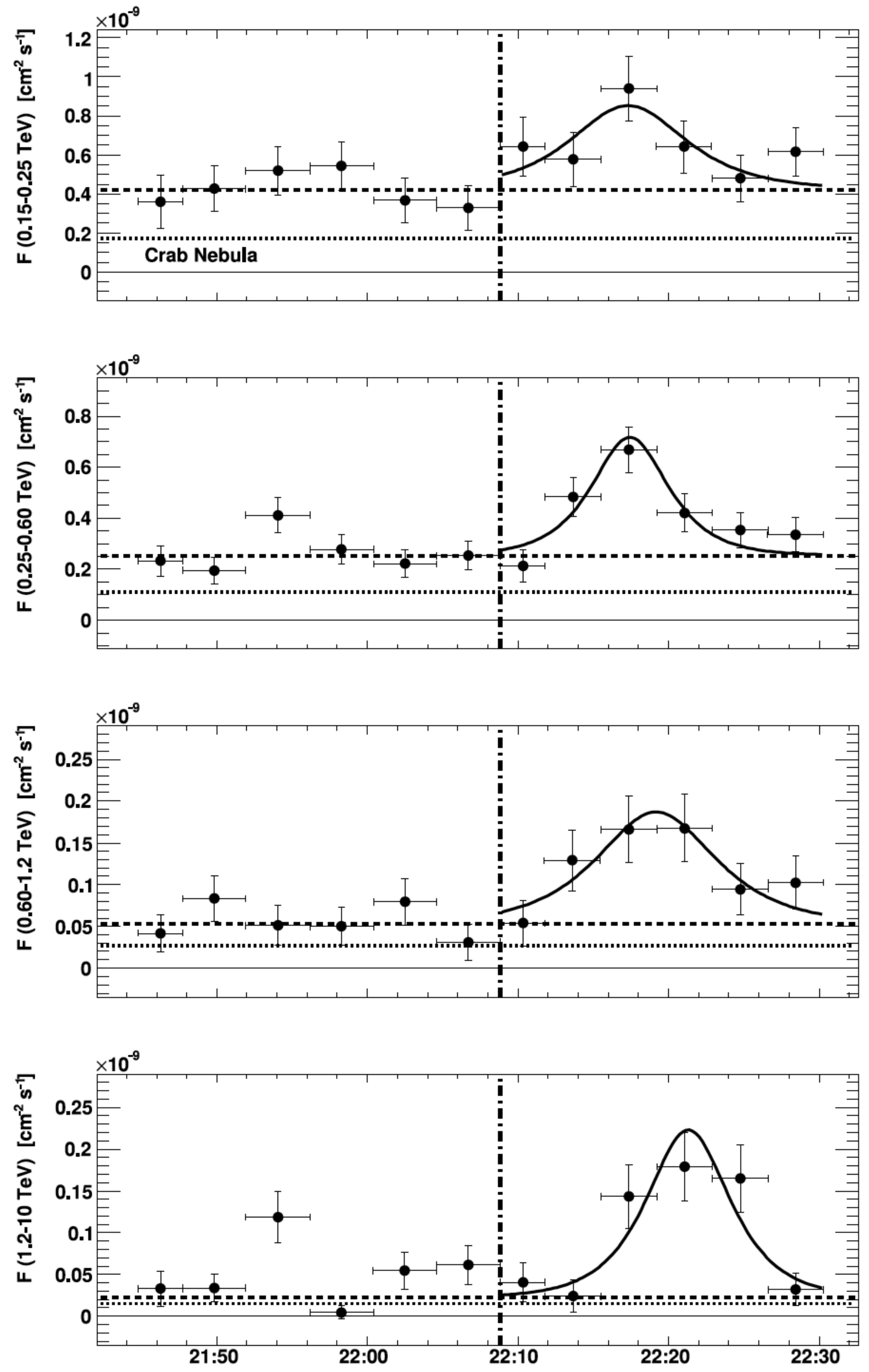}
\caption{Gamma-ray light curve during the night of July 9 2005, with
  tim binning of 4 minutes and in different energy bands. The dotted
  line represents the flux from the Crab Nebula. Figure
  taken from \cite{MrkMagic}.}
\label{fig:MrkMagic}
\end{figure}

The effect is expected to be tiny for energies below $M_P$, but
nevertheless measurable in the time delay of photons traveling
cosmological distances (from far AGNs or GRBs), observed over a
wide-enough energy range. The time delay per unit energy for gamma
rays traveling a distance $z$ can be written as
\begin{equation}
\frac{\Delta t}{\Delta E} \simeq \frac{\xi}{M_PH_0} \int_0^z
\frac{(1+z')dz'}{\sqrt{\Omega_m(1+z')^3+\Omega_\Lambda}}
\label{eq:linear}
\end{equation}
Or, if $\xi=0$, the time delay per unit of energy \emph{square} can be
written as:
\begin{equation}
\frac{\Delta t}{\Delta E^2} \simeq \frac{3\zeta^2}{2M_P^2H_0} \int_0^z
\frac{(1+z')^2dz'}{\sqrt{\Omega_m(1+z')^3+\Omega_\Lambda}}
\label{eq:quadratic}
\end{equation}

The first limits were obtained using MAGIC observations
(\cite{LIVMagic}) of an intense flare in Markarian 501
(\cite{MrkMagic}), a relatively close AGN ($z = 0.034$). Variations in
the flux of factors close to 10 were observed, with doubling times of
the order of 1-2 minutes, in the energy range between 150 GeV and 10
TeV (see Figure~\ref{fig:MrkMagic}). Several innovative analysis
methods were specially developed for this study, and the measured time
delays per unit energy and energy \emph{square} were, respectively:
$\tau_l = (30\pm 12)$ s/TeV and $\tau_q = (3.7 \pm 2.6)$
s/TeV$^2$. Although these results could be interpreted as significant
detection of an energy-dependence of the speed of gamma rays, delays
produced at the source cannot be excluded by a single observation. The
corresponding 95\% confidence level (CL) lower limits to the mass
scale of Lorentz Invariance violation are $M_P/\xi > 0.3 \times
10^{18}$ GeV and $M_P/\zeta > 5.7 \times 10^{10}$ GeV.

 Shortly after, H.E.S.S. observed an exceptionally intense flare in PKS
2155-304 ($z=0.116$) happened in July 28, 2006, with flux peaks up to
15 times the Crab flux and doubling times of 1-3 minutes. The energy
range of the observations covered from 200 GeV to 4 TeV. The measured
time delays were $\tau_l = (-6 \pm 11)$ s/TeV and $\tau_q = (1.7 \pm
6.3)$ s/TeV$^2$, corresponding to limits to the mass scale of Lorentz
Invariance violation: $M_P/\xi > 2.1 \times 10^{18}$ GeV and
$M_P/\zeta > 6.4 \times 10^{10}$ GeV (\cite{LIVHess}).

Fermi-LAT can observe sources from the farthest distances in a
comparatively narrower energy range, of the order of tens of
GeV. The observation of the GRB on May 10th 2009 ($z=0.903$) in
the energy range between 10 MeV and 30 GeV (\cite{LIVFermi}), whose
duration was shorter than 1 s, allowed to set an upper limit to the
gamma-ray time delay (considering the linear term in
Equation~\ref{eq:linear}) of $\tau_l < 30$ s/TeV. The corresponding
lower limits to the mass scale of Lorentz Invariance violation are
$M_P/\xi > 1.5 \times 10^{19}$ GeV and $M_P/\zeta > 3.0 \times
10^{10}$ GeV.

Fermi-LAT constrains better the linear term (see
Equation~\ref{eq:linear}), due to the larger distances traveled by
the observed gamma rays and the limit exceeds, for the first time in this
kind of tests, the Planck Mass. The quadratic term is better
constrained by observations with IACTs, due to the wider accessible
energy interval. For a summary of the limits obtained with the
different observatories, see Table~\ref{table:liv}.

\begin{table}[t]
\begin{center}
\caption{95\% CL lower limits on the mass scale of Lorentz Invariance violation
  obtained from gamma-ray observations}
\begin{tabular}{|l|c|c|}
\hline \textbf{Telescope} & \textbf{$M_P/\xi$} [GeV]&
\textbf{$M_P/\zeta$} [GeV] 
\\
\hline 
MAGIC       & $0.03 \times 10^{19}$ & $5.7 \times 10^{10}$ \\
H.E.S.S.      & $0.21 \times 10^{19}$ & $6.4 \times 10^{10}$ \\
Fermi-LAT & $1.50 \times 10^{19}$ & $3.0 \times 10^{10}$ \\
\hline
\end{tabular}
\label{table:liv}
\end{center}
\end{table}

\section{DARK MATTER SEARCHES}
\label{sec:dm}

In indirect DM searches with gamma rays we look for either primary or
secondary gamma rays produced in annihilation and/or decay of DM
particles. In first approximation, we can consider that these gamma
rays do not interact on their way to Earth, since all relevant known
source candidates are at relatively short --non-cosmological--
distances. In addition, having no electric charge, gamma rays arrive
from the direction of the annihilation/decay site, providing clear
spatial signatures. For the case of DM annihilation, the expected
gamma-ray flux from a region of the sky $\Omega$ can be written as
follows:
\begin{equation}
\frac{d\Phi}{dE}(\Omega) = \frac{\langle \sigma
  v\rangle}{2m_\textrm{dm}} \frac{dN_\gamma}{dE}\,  \frac{1}{4\pi} \int_\Omega \int_\textrm{los}
\rho^2(l)\, dl\, d\Omega'  
\label{eq:dm}
\end{equation}
where $\langle \sigma v\rangle$ is the averaged annihilation
cross section times velocity of the DM particle, $m_\textrm{dm}$ the
mass of the DM particle, $N_\gamma$ is the number of gamma rays
produced per annihilation process, $E$ the energy of the gamma ray, and
$\rho$ the DM density. The integral runs over the line of sight ($l$)
and the observed sky region ($\Omega$).

\begin{figure}
\includegraphics[width=80mm]{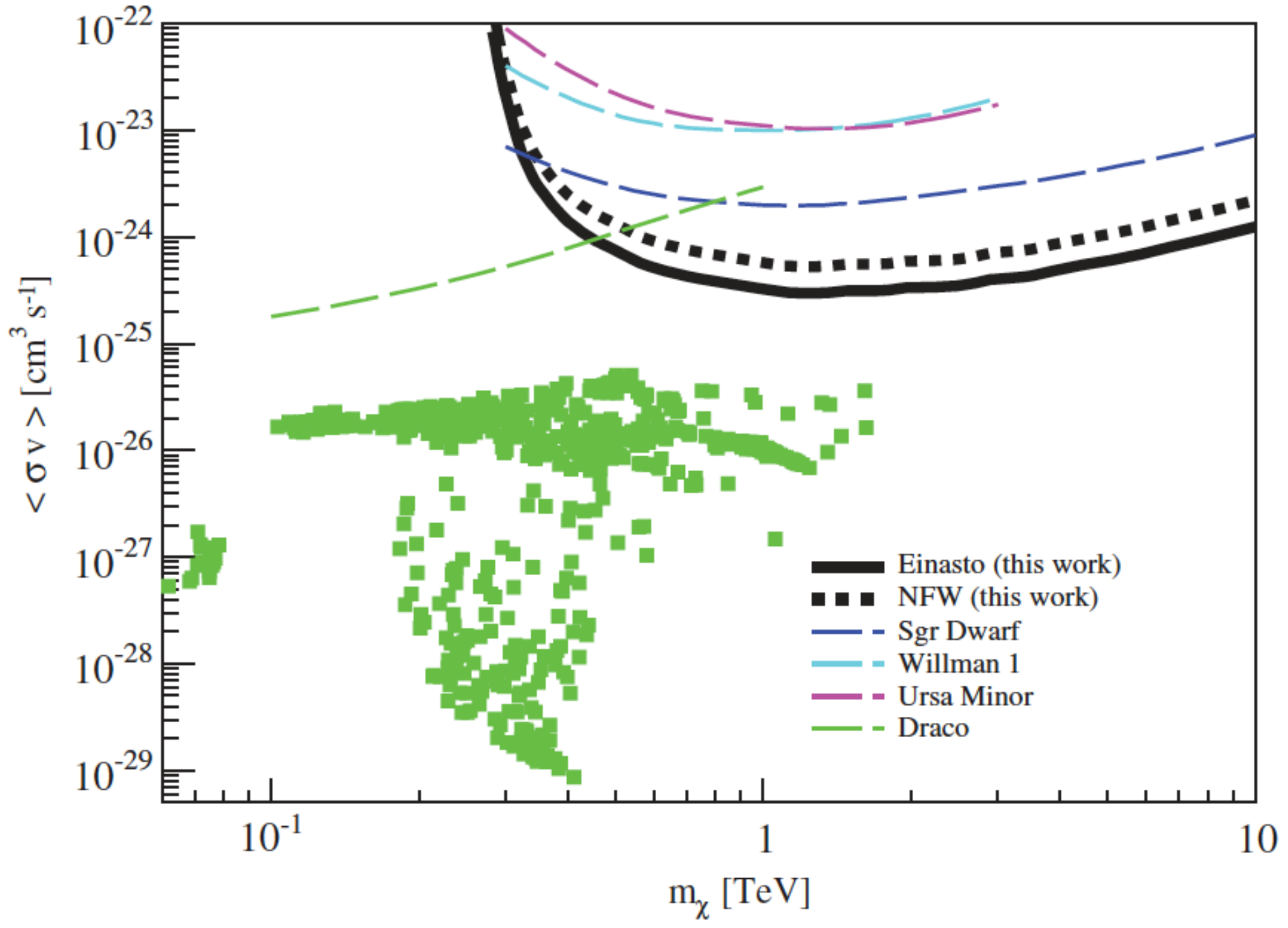}
\caption{Upper limits to  $\langle\sigma
v\rangle$ (in the mSUGRA scenario) obtained by observations of
H.E.S.S. on the Galactic halo. Figure
  taken from \cite{dmHess}.}
\label{fig:dmHess}
\end{figure}

The first part of Equation~\ref{eq:dm} contains all the information
dependent on the specific nature of the DM particle and its
interactions. $\frac{dN_\gamma}{dE}$ contains all the spectral
information, and provides unique, universal spectral
features. In a real experiment we observe $\frac{dN_\gamma}{dE}$
convoluted with the energy response function of the instrument,
containing the detection efficiency, the energy resolution and
bias. On the other hand, the integral in Equation~\ref{eq:dm}
determines the total flux normalization, and contains all the
information regarding the source distance and geometry. It depends on
the specific target of the observation (e.g.\ the Galactic center or
halo, satellite galaxies or galaxy clusters), and is affected by large
theoretical uncertainties from halo simulations. Moreover, in real
experimental conditions we observe it convoluted with the point-spread
function of the instrument, which, in general, degrades the
sensitivity, particularly for point-like signals.

\begin{figure}
\includegraphics[width=80mm]{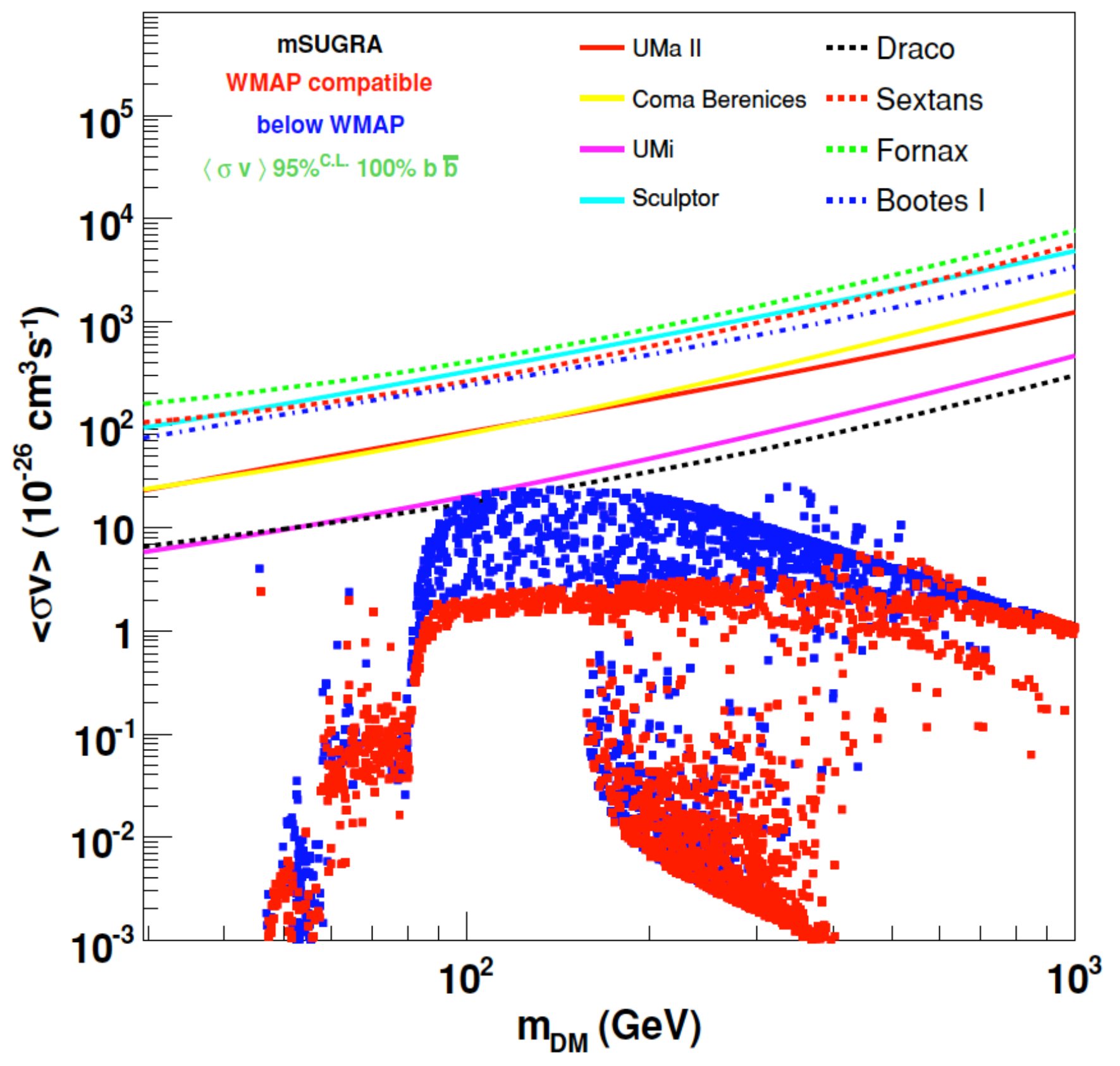}
\caption{Upper limits to  $\langle\sigma
v\rangle$ (in the mSUGRA scenario) obtained by observations of
Fermi-LAT on dwarf satellite galaxies. Figure taken from \cite{dmFermi}.}
\label{fig:dmFermi}
\end{figure}

All gamma-ray instruments have observed various of the most promising
candidate DM-induced sources. The most stringent limits obtained by
MAGIC come from 30 hours of observations of the satellite galaxy Segue
1 (\cite{dmMagic}), reaching limits of the order of $\langle\sigma
v\rangle \sim\times 10^{-23}$ cm$^3$ s$^{-1}$ for DM masses
$\sim$few$\times 100$ GeV (considering mSUGRA models). Similar limits
are also obtained using observations of satellite galaxies with
VERITAS (\cite{dmVeritas}). 

The most stringent limits obtained with IACTs are provided by H.E.S.S.,
using observations of the Galactic halo (\cite{dmHess}), which yield
limits at the level of $\langle\sigma v\rangle \sim$ few$\times
10^{-25}$ cm$^3$ s$^{-1}$ for a similar DM range (see
Figure~\ref{fig:dmHess}).

For DM mass below a few$\times$100 GeV, the best limits are provided
by the observations of Fermi-LAT. Thanks to its large FoV and duty
cycle, Fermi observes all possible candidates almost
simultaneously. Observations of dwarf satellite galaxies
(\cite{dmFermi,dmFermi2}) provide limits maybe reaching
$\langle\sigma v\rangle \sim\times 10^{-25}$ cm$^3$ s$^{-1}$ for a DM
mass $\sim 100$ GeV (see Figure~\ref{fig:dmFermi}).

The ultimate sensitivity in DM searches using the present generation of
gamma-ray instruments has not been still reached. Deeper
observations, together with improvements in the analysis techniques
will produce in the next few years a more sensitive search, reaching
predictions of mSUGRA, and will be continued in the future by
CTA.

\section{SUMMARY AND CONCLUSIONS}
\label{sec:conclusions}

In this paper we have shown how gamma-ray instruments have been (and
still are) used to probe several topics of fundamental physics. We
have summarized the most relevant results in several fronts: \emph{i)}
Gamma-ray instruments are gathering evidence on the role of SNRs as
the primordial sites of CR acceleration. \emph{ii)} CR
electrons-positrons have been measured between 30 GeV and 5 TeV,
confirming a harder spectrum than previously thought but discarding
sharp spectral features. \emph{iii)} Models of EBL have been
constrained down to almost the lower limits allowed by galaxy
counts. \emph{iv)} The Quantum Gravity scale has been probed up to Planck
Mass in Lorentz Invariance tests \emph{v)} Ongoing DM searches have
not yielded yet any positive result, but are expected to constrain
mSUGRA models in the coming years.

\bigskip % extra skip inserted
\begin{acknowledgments}
  I thank the MAGIC, H.E.S.S., VERITAS and Fermi-LAT Collaborations
  for their help in reviewing and improving this
document.\end{acknowledgments} \bigskip % extra skip inserted
%% Create the reference section using BibTeX:
\bibliography{Rico_GammaRay}

\end{document}